# Interaction of Tannin with Bovine Serum Albumin by Fluorescence Spectrometry

Kim Dong-Il[1], Choe Son-Ae[1], Sin Kye-Ryong[2,*]

([1]Institute of Analysis, **Kim Il Sung** University, Pyongyang, DPR Korea,

[2]Faculty of Chemistry, **Kim Il Sung** University, Pyongyang, DPR Korea)

* E-mail : ryongnam9@yahoo.com

**Abstract:**

Interaction between tannin and bovine serum albumin (BSA) was examined by the fluorescent quenching. The process of elimination between BSA and tannin was the one of a stationary state, and the coupling coefficient was one, the coupling constant was $2.89 \times 10^5$ $(mol/L)^{-1}$. The working strength between the tannin and the beef serum was hydrophobic one.

## Introduction

Tannin, the derivative of polyphenol spreaded abundantly in the floral kingdom is widely used as an adstringentia, an antidote and a tanning agent.[1-5]

After entering into the blood, drugs can be distributed the whole body through blood circulation and occurred the pharmacological action. In this course, medical molecules surely combine in different degrees with serum proteins (mainly serum albumin). This binding reaction has an influence on the acting time and strength of the drug.

Therefore, study on the interaction of the drug molecule with serum albumin is very significant in the drug thermodynamics and the clinical pharmacology.

There are many reports on the interactions of drugs with human serum albumin (HSA) and BSA, but it is difficult to find out the papers on the interaction of tannin with BSA by fluorometry.[5-8]

In this paper, the binding constants, the number of binding sites and the binding force between the tannin and the BSA were determined by applying the fluorescence quenching method to the interaction of tannin with BSA.

## Instrumentations and Reagents

Here used were A RF-5000 Spectrofluorophotometer equipped with a 150w xenon arc lamp for excitation and UV-2201 ultraviolet visible spectrophotometer using a quartz cell.

All reagents used were of AR grade. Tannin, BSA and phosphate buffer solution (pH 7.4) were used. All aqueous solutions were prepared in doubly deionized water.

## Experimental

The experiment procedures were performed as follows: an appropriate quantity of BSA solution and tannin solution were transferred into a 10 ㎖ volumetric flask and phosphate buffer solution (pH 7.4) was added until the scale and mixed thoroughly mixture.

The solution was stabilized in the thermostat for 10 minutes and then was measured by fluorometer and UV-Vis spectrometer.

**Results and Discussion**

In the excitation wavelength 280nm, BSA exhibits the fluorescence in the wavelength range of 300 to 500nm, by the Trp and Tyr in the protein, but the tannin does not.

While the concentration of BSA was fixed and the one of tannin was gradually increased, synchronous fluorescence spectrum was measured in $\Delta\lambda=80$nm (Fig 1).

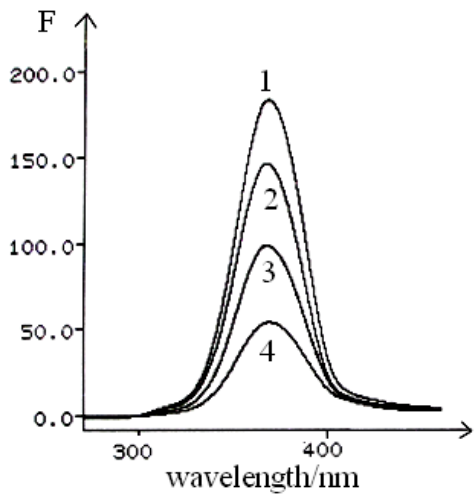

Fig 1. Synchronous fluorescence spectra of BSA in the presence of Tannin
(BSA:$1\times10^{-6}$mol/L, 1~4: Tannin: (0, 0.2, 0.4, 0.6) $\times 10^{-6}$mol/L, at 298.15K)

As can be seen from Fig 1, there was no change of the wavelengths of fluorescence peaks, whereas fluorescence intensities of BSA decrease according to the increase of tannin's concentration. In the meantime, tannin has no fluorescence emission at range of this wavelengths. This shows that there were interactions between tannin and BSA.

The fluorescence quenching is very important to find out the interaction between the protein and the quencher. There are the various fluorescence quenching, such as a static, dynamic and energy transfer quenching.

To determine the type of fluorescence quenching, the absorption spectra of BSA and the difference of absorption spectra between BSA-tannin compound and tannin were measured. (Fig 2)

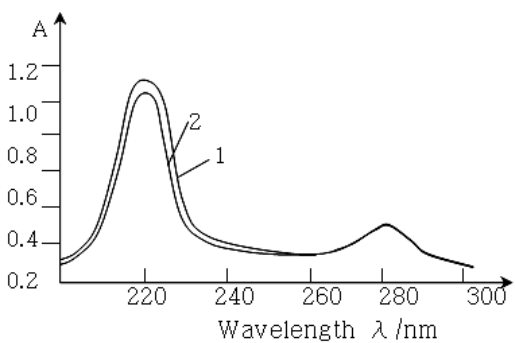

Fig 2. Influence of tannin on the absorption spectra intensity of BSA.
(1: Absorption spectrum of BSA. 2: Difference of absorption spectrum between the BSA-tannin compound and tannin. $C_{BSA}$:$1\times10^{-6}$mol/L, $C_{tan}$: $1\times10^{-6}$mol/L)

As can be seen from Fig 2, BSA has maximum absorbance peak at 220nm and the absorbance of BSA decreased with the addition of tannin. This shows that BSA has an interaction with tannin at the ground state and makes a form of coordination compound, thus, the process of decreasing the fluorescence intensity of fluorescent material is the static quenching one. Dynamic quenching has an influence on the excited state of fluorescent material, but does not change its absorbance spectra.

The graph according to Stern-Volmer equation was drown by measuring fluorescence spectra of BSA-tannin compound at various temperatures while the concentration of BSA was fixed and the concentration of tannin was gradually increasing. (Fig 3)

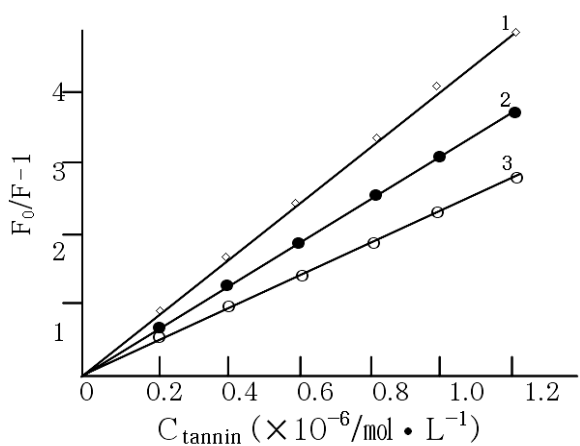

Fig 3. Graphs according to Stern-Volmer equation of BSA-tannin compound at various temperatures. ($C_{BSA}$:1×10$^{-6}$mol/L, 1-20℃, 2-30℃, 3-45℃)

As can be seen from Fig 3, the gradient of plot decreased according to the increase of temperature. If fluorescence quenching is dynamic, the gradient of plot equal to product, $K_q \cdot \tau_0$ of the quenching rate constant, $K_q$, and the fluorescence lifetime, $\tau_0$, in absence of quencher. The fluorescence lifetime of organic macromolecule is nearly $10^{-8}$s.

The determined quenching rate constants were showed in Table 1, where the quenching rate constants was bigger than the maximum quenching constant of diffuse collision between various kinds of quenchers and organic macromolecule, $2.0 \times 10^{10}$ L/(mol·s), and the one at lower temperature is bigger than that at higher temperature. This indicates that fluorescence quenching process of the BSA-tannin compound was the static one.

Quenching rate constants / L·mol$^{-1}$·s$^{-1}$    Table1

| Temperature/℃ | $K_q$ | Correlation coefficient |
|---|---|---|
| 20 | 3.68×10$^{13}$ | 0.995 |
| 30 | 3.58×10$^{13}$ | 0.996 |
| 45 | 3.35×10$^{13}$ | 0.995 |

As the fluorescence quenching process was static one, we draw the graph according to the static fluorescence quenching equation lg(($F_0$-F)/F)=lgK+nlgC$_{tan}$ (F and $F_0$ is the relative fluorescence intensity in presence and in absence of tannin and n the number of binding site and K binding constant) and n, the number of binding sites calculated by the gradient of the plot, was approximately 1. (Fig 4, Table 2)

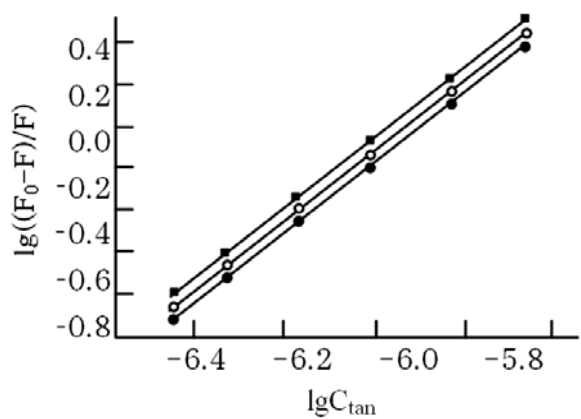

Fig 4. Plot of $\lg((F_0-F)/F)$ vis $\lg C_{tan}$ at different temperature
($C_{BSA}$:1×10$^{-6}$mol/L, 1: 20℃, 2: 30℃, 3: 45℃)

Meanwhile, the number of binding sites was determined by molar ratio method on the basis of tannin's quenching action. (Fig 5)

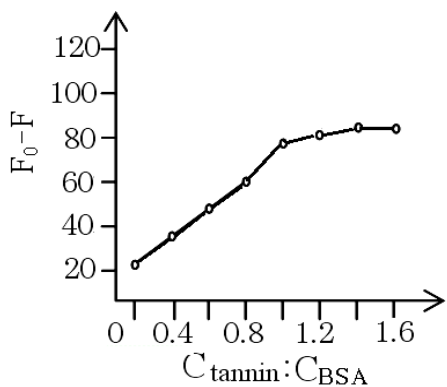

Fig 5. Fluorescence quenching efficiency of BSA according to the concentration of tannin

As can be seen from Fig 5, when the ratio of the concentration of tannin ($C_{tan}$) and BSA ($C_{BSA}$) was more than 1, the binding was saturated. This shows that tannin was bound with BSA with the ratio of 1:1, that means the number (n) of binding sites of tannin on BSA is 1. The binding constant K was determined by using the intercept value on Fig 4 (Table 2).

The binding constant and the number of binding sites    Table 2

| Temperature/℃ | K/(mol·L$^{-1}$) | n |
|---|---|---|
| 20 | 2.89×10$^5$ | 1.16 |
| 30 | 2.53×10$^5$ | 1.07 |
| 45 | 2.16×10$^5$ | 1.04 |

From table 2, it can be seen that the binding constant was decreasing while the temperature was getting higher. The binding constant of interaction between the tannin and BSA was comparatively big, therefore it can be used for the storage and the transport of the protein in body as there was rather strong binding force.

The binding force of protein with drugs was different according to the various drug molecules. The interaction between protein and drugs belong to weak interaction between molecules. If ΔH>0 and ΔS>0, then it is the hydrophobic interaction, if ΔH<0 and S>0, then the electrostatic, and if ΔH<0 and S<0, then the Van der waals and the hydrogen-bonding.[11]

To determine the sort of binding force, we determine ΔG (ΔG=-RTlnK) of binding reaction from the binding constant between BSA and tannin and ΔH (ΔH=(1/$T_1$-1/$T_2$)-1·R·ln($K_1$/$K_2$) ) of the binding reaction from the binding constant $K_1$ and $K_2$ at temperature $T_1$ and $T_2$ and ΔS (ΔS=(ΔH-ΔG)/T) of binding reaction. The obtained ΔH and ΔS was 16.7kJ/mol, 148.06J/(mol·K), respectively and both were biger than 0, so this indicates the interaction between tannin and BSA was mainly the hydrophobic interaction.

**Conclusion**

Here discussed was the quenching action of tannin on fluorescence of BSA and found that it was static. The BSA and tannin was bound each other mainly due to the hydrophobic interaction force and the number of binding site of tannin in range located on the BSA tryptophan residue was 1 and the binding constant was 2.89×10$^5$mol/L. There was no remarkable change on the structure of BSA when the tannin was added.

**References**
[1] Tian-ying Zhang, Gang Li, Hai-zhen Mo and Chun-xiang Zhi, Persimmon tannin composition and function, Advances in Biomedical Engineering, 1-2, 389, 2011.
[2] Chun-mei Li, R.L.John, D. Trombley, Shu-fen Xu, J. Yang, and Y. Tian, High molecular weight persimmon (Diospyros kaki L.) proanthocyanidin: a highly galloylated, a-linked tannin with an unusual flavonol terminal unit, myricetin. J. Agri. Food Chem., 58, 9033–9042, 2010.
[3] I. Tarascou, J. P. Mazauric, S. Carrillo, S. Coq, F. Canon, The hidden face of food phenolic composition. Archives of Biochemistry and Biophysics, 501(1), 16-22, 2010.
[4] Akagi T., Ikegami A, et al., Condensed tannin composition analysis in persimmon (Diospyros kaki Thunb.) fruit by acid catalysis in the presence of excess phloroglucinol, J. Japan. Soc. Hort. Sci., 79(3), 275-281, 2010.
[5] Kawakami K, Nakanami M, Iizuka S, Hirayama M., Major-water-soluble polyphenols, proanthocyanidins in leaves of persimmon (Diospyros Kaki) and their α-amylase inhibitor y acitivity. Biosci. Biotechnol. Biochem, 74(7), 1380-1385. 2010.
[6] He Wen-ying, Chen Guang-ying, Du.Juan, Yao Xiao-jun, Investigation on the binding of piperine to bovine serum albumin by optical spectroscopy and molecular modeling, Acta Chimica Sinica, 66(21), 2365, 2008.
[7] He Wen-ying, Chen Guang-ying, Zhang Lian-hua, Zhang Zheng, Study on the interaction of vinblastine sulfate and bovine serum albumin, J. Analytical Science, 24(6), 659, 2008.
[8] Cui Feng-ling, Cui Yan-rui, Luo Hong-xia, Yao Xiao-jun, Fan Jing and Lu Yan, Interaction of APT with BSA or HAS, Chinese Science Bulletin, 51(18), 2201, 2006.